\documentclass[twocolumn,showpacs,prb,amsmath,amssymb,superscriptaddress]{revtex4-1}

\usepackage{color}
\usepackage{epstopdf}
\usepackage{graphicx}
\usepackage{dcolumn}
\usepackage{bm}

\begin{document}

\preprint{APS/123-QED}

\title{Fermi surfaces and orbital polarization in superconducting
 CeO$_{\bm{0.5}}$F$_{\bm{0.5}}$BiS$_{\bm{2}}$ \\
revealed by angle-resolved photoemission spectroscopy}

\author{T. Sugimoto}
\author{D. Ootsuki}
\affiliation{Department of Physics \&  Complexity Science and Engineering, University of Tokyo, 5-1-5 Kashiwanoha 277-8561, Japan}

\author{C. Morice}
\affiliation{Cavendish Laboratory, University of Cambridge, Cambridge CB3 0HE, United Kingdom}

\author{E. Artacho}
\affiliation{Cavendish Laboratory, University of Cambridge, Cambridge CB3 0HE, United Kingdom}
\affiliation{Nanogune and DIPC, Tolosa Hiribidea 76, 20018 San Sebasti\'{a}n, Spain}
\affiliation{Basque Foundation for Science, Ikerbasque, 48011 Bilbao, Spain}

\author{S. S. Saxena}
\affiliation{Cavendish Laboratory, University of Cambridge, Cambridge CB3 0HE, United Kingdom}

\author{E. F. Schwier}
\affiliation{Hiroshima Synchrotron Radiation Center, Hiroshima University, Higashihiroshima 739-0046, Japan}

\author{M. Zheng}
\author{Y. Kojima}
\affiliation{Graduate School of Science, Hiroshima University, Higashihiroshima 739-8526, Japan}

\author{H. Iwasawa}
\author{K. Shimada}
\author{M. Arita}
\author{H. Namatame}
\affiliation{Hiroshima Synchrotron Radiation Center, Hiroshima University, Higashihiroshima 739-0046, Japan}

\author{M. Taniguchi}
\affiliation{Hiroshima Synchrotron Radiation Center, Hiroshima University, Higashihiroshima 739-0046, Japan}
\affiliation{Graduate School of Science, Hiroshima University, Higashihiroshima 739-8526, Japan}

\author{M. Takahashi}
\affiliation{Department of Physics \&  Complexity Science and Engineering, University of Tokyo, 5-1-5 Kashiwanoha 277-8561, Japan}

\author{N.~L.~Saini}
\affiliation{Dipartimento di Fisica, Universit\'a di Roma ``La Sapienza'' - Piazzale Aldo Moro 2, 00185 Roma, Italy}

\author{T. Asano}
\author{T. Nakajima}
\author{R. Higashinaka}
\author{T. D. Matsuda}
\author{Y. Aoki}
\affiliation{Department of Physics, Tokyo Metropolitan University, Hachioji 192-0397, Japan}

\author{T. Mizokawa}
\affiliation{Department of Applied Physics, Waseda University, Tokyo 169-8555, Japan}

\date{\today}

\begin{abstract}
We have investigated the electronic structure of BiS$_2$-based CeO$_{0.5}$F$_{0.5}$BiS$_2$ superconductor using polarization-dependent angle-resolved photoemission spectroscopy (ARPES), and succeeded in elucidating the orbital characters on the Fermi surfaces. In the rectangular Fermi pockets around X point, the straight portion parallel to the $k_y$ direction is dominated by Bi $6p_x$ character. The orbital polarization indicates the underlying quasi-one-dimensional electronic structure of
the BiS$_2$ system. Moreover, distortions on tetragonally aligned Bi could give rise to the band Jahn-Teller effect.
\end{abstract}

\maketitle

\newpage

Since the discovery of BiS$_2$-based superconductors by Mizuguchi {\it et al.} \cite{Mizuguchi2012} in 2012, various BiS$_2$-based systems have been discovered including RE(O,F)BiS$_2$ system (RE=rare earth element). In a typical RE(O,F)BiS$_2$ system, electronically active BiS$_2$ layers (or BiS planes) are sandwiched by RE(O,F) layers, and the F substitution is believed to be electron-doping to the electronically active BiS$_2$ layer. In the various RE(O,F)BiS$_2$ systems, numerous theoretical and experimental researches have been performed in order to understand their electronic states including the pairing mechanism \cite{2,3,4,5,6,7,8,9,10,11,12,13,14,15,16,Demura2015,Sugimoto2014,Sugimoto2015}.
Among them, the Fermi surfaces and the band dispersions of RE(O,F)BiS$_2$ 
have been investigated by means of angle-resolved photoemission spectroscopy (ARPES) \cite{Ye2014,Saini2014,Zeng2014}. 
The observed non-dispersive $k_x$-$k_z$ Fermi surfaces \cite{Ye2014} due to the two-dimensional-like crystal structure are consistent with the decoupled BiS plane with block layer by the F-doping reported by local structural studies \cite{Paris2014}. This indicates the absence of Bi $6p_z$ electrons near Fermi level ($E_F$).
The observed band structure and Fermi surface geometry are also basically consistent with the band structure calculations, except for that the area of the Fermi surface is not fully reproduced. For instance, in the optimal F-doped region ($x=$0.18), the total area of the Fermi surfaces is consistent with the amount of F substitution  in a similar selenide superconductor La(O,F)BiSe$_2$ system\cite{Saini2014}. However, in the higher doped region ($x\sim0.5$),  
the total area of the observed Fermi surfaces was much smaller than that expected from
the nominal amount of F substitution in Nd(O,F)BiS$_2$ \cite{Zeng2014,Ye2014}.
The discrepancy in the Nd(O,F)BiS$_2$ system suggests that a part of electrons introduced by F substitution would be localized and do not contribute to the Fermi surfaces.


According to the theoretical studies, in a typical RE(O,F)BiS$_2$ system, all the Bi $6p$ orbitals have almost the same band center ($E=E_F + 2.9$ eV), and the Fermi surfaces are constructed from the  $p_x$ and $p_y$ orbitals due to their much wider band widths than that of $p_z$ \cite{Morice2013,Usui2012}. 
Near X point where Fermi surfaces exist, the $p_x$ and $p_y$ orbitals may not hybridize and form the straight Fermi surfaces along the $k_y$ and $k_x$ directions, respectively. In this situation, the BiS$_2$ system can be viewed as a quasi-one-dimensional electron system \cite{Usui2012} despite its two-dimensional-like layered crystal structure. The quasi-one-dimensional character of the electronic states tends to enhance the Peierls instability due to electron-lattice interaction and/or or the charge fluctuation due to electron-electron interaction and may cause the deviation from the band structures obtained by the local density approximation (LDA) \cite{Mizokawa2005}.

In this paper, in order to address these issues mentioned above (smaller Fermi surface, polaron, quasi-one-dimensionality), we have performed polarization-dependent high resolution ARPES measurement and band-structure calculation on CeO$_{0.5}$F$_{0.5}$BiS$_2$.


\begin{figure}
\includegraphics[width=8cm]{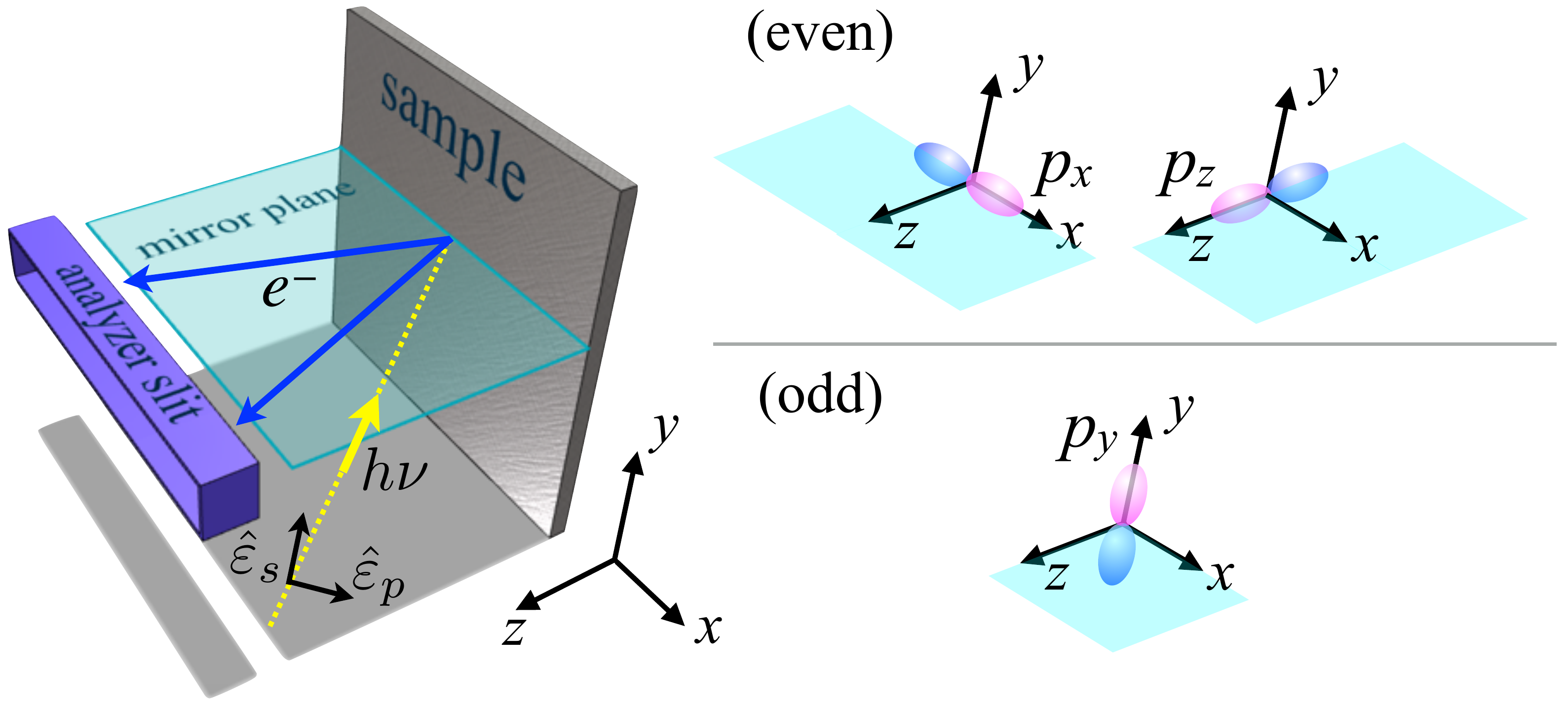}
\caption{(Color online) Schematic diagrams of measurement geometries with linearly polarized photons that give us $p$-polarization and $s$-polarization.  The corresponding polarization vectors are denoted by $\hat{\varepsilon}_p$ and $\hat{\varepsilon}_s$.The $p$-orbitals with even and odd symmetries with respect to the mirror plane are also shown in the right.  
}
\label{fig1}
\end{figure}

High-quality single crystal of CeO$_{0.5}$F$_{0.5}$BiS$_2$ has been prepared by CsCl flux method, and its details are given elsewhere \cite{Higashinaka2015}.
The ARPES measurements with linearly/circularly polarized photons were performed at the undulator beamlines BL-1/BL-9A of the Hiroshima Synchrotron Radiation Center, Hiroshima University. Both endstations are equipped with VG SCIENTA R4000 analyzer. The photon energy was set to be 30 eV for all the polarizations. We cleaved the single crystalline sample \textit{in situ} in ultrahigh vacuum ($<1\times10^{-10}$ Torr) in order to obtain clean (001) surface. The total energy resolution including both monochromator and electron energy analyzer was measured to be 21 meV.  All the measurements were taken at 50 K. The experimental $p$- and $s$-polarization setups are shown in Figs. \ref{fig1}. In photoemission process, the transition matrix element $M_{\text{i,f}}^{\bm{k}}$ can be described as
\begin{equation}
|M_{\text{i,f}}^{\bm{k}}| \propto ||\langle \phi^{\bm{k}}_{\text{f}}|\hat{\varepsilon}\cdot \bm{r} |\phi^{\bm{k}}_{\text{i}} \rangle||^2 
\label{eq1}
\end{equation}
where $ \phi^{\bm{k}}_{\text{i}}$ ($\phi^{\bm{k}}_{\text{f}}$) is the initial- (final-) state wave function, and $\hat{\varepsilon}$ is the polarization vector of the incident photons. The final state $\phi^{\bm{k}}_{\text{f}}$  is always even since it is in the mirror plane and can be approximately described by a plane wave. In the $p$- ($s$-) polarization setup, the term $\hat{\varepsilon}\cdot \bm{r}$ is even (odd), and the corresponding initial state should be even (odd) with respect to the mirror plane in order to give non-vanishing $M_{\text{i,f}}^{\bm{k}}$. Therefore, the $p_x$ and $p_z$ orbitals are observable for the $p$-polarization setup, while the $p_y$ orbital for the $s$-polarization setup, as depicted in Fig. 1. Unlike the linear polarization, the dipole selection rule does not hold for circularly polarized photons giving non-zero $M_{\text{i,f}}^{\bm{k}}$ for all the orbitals.

Band structures were calculated using the SIESTA method \cite{Artacho2008,Soler2002}, implementing the generalized gradient approximation in the shape of the Perdew, Burke, and Ernzerhof functional\cite{Perdew1996}. It uses norm-conserving pseudopotentials to replace the core electrons, while the valence electrons are described using atomic-like orbitals as basis states at the double zeta polarized level. We performed GGA+$U$ calculations \cite{Anisimov1991} because of the known strong correlation of the 4$f$ electrons in cerium. As no atomic coordinates have been reported yet, we obtained a geometry from density functional theory, by relaxing the structure (both atoms and cell), starting from the structure known for LaO$_{0.5}$F$_{0.5}$BiS$_{2}$. The band structure calculations are further detailed in Ref. 25.


\begin{figure}
\includegraphics[width=8.5cm]{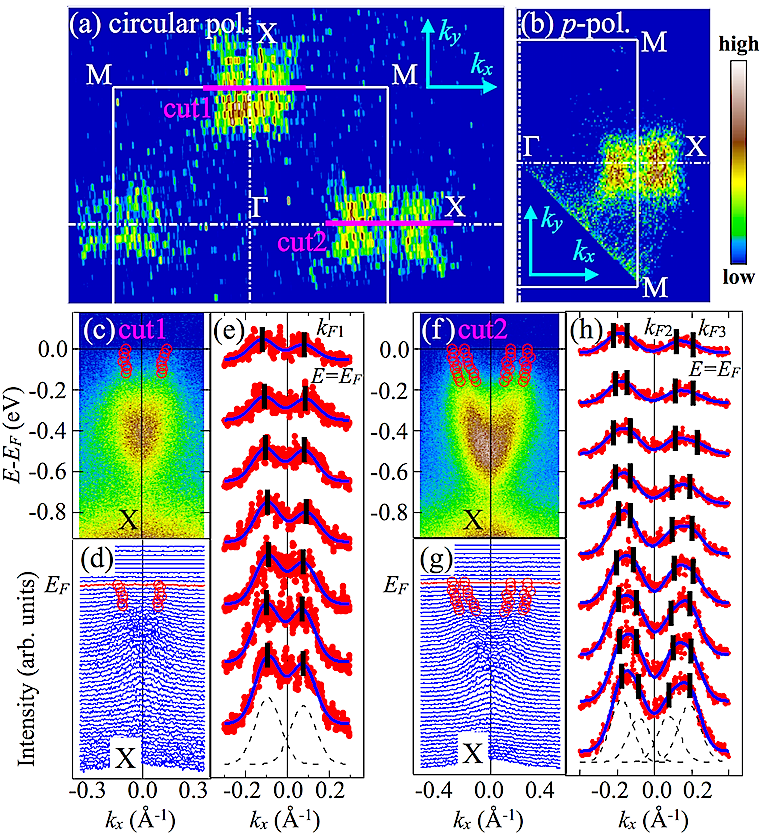}
\caption{(Color online) (a) Fermi surface maps of CeO$_{0.5}$F$_{0.5}$BiS$_2$ taken with circularly polarized light and (b) with $p$-polarization geometry, integrated within $\pm$40 meV.  (c) Cut denoted by `cut1' in (a). (d) MDCs, (e) fit on MDCs from $E=-0.12$ eV up to $E_F$ for the extraction of peak positions.  (f) Cut denoted by `cut2' in (a). (g) MDCs, (h) fit on MDCs from $E=-0.16$ eV up to $E_F$ for the extraction of peak positions. Circles in (c,d,f,g) are the peak positions of MDCs, dots (solid lines, bars) in (e,h) are the experimental MDCs (fit, peak positions) near $E_F$. Gaussian functions giving the best fit to the bottommost MDCs are also displayed in (e,h).  All the MDCs were obtained with 20 meV energy step.
 }
\label{fig2}
\end{figure}

The Fermi surface map of CeO$_{0.5}$F$_{0.5}$BiS$_2$ was taken by changing tilt angle with circularly polarized light as shown in Fig. \ref{fig2}(a) whereas that with the $p$-polarized light was taken by changing the azimuthal angle as shown in Fig. \ref{fig2}(b). Intensities of the Fermi surface maps were integrated within $\pm$40 meV above and below $E_F$.
In the rectangular Fermi pockets with $p$-polarized light [Fig. 2(b)], the straight portion parallel to the $k_y$ direction is much more enhanced than that parallel to the $k_x$ direction. 
This observation may indicate that the straight portion parallel to the $k_y$ direction is derived from Bi 6$p_x$ and that the other straight portion parallel to the $k_x$ direction has more Bi 6$p_y$ character. However, since the Fermi surface was measured by changing the azimuthal angle, the orbital selectivity is not exact for the straight portion parallel to the $k_x$ direction.
Figures 2(c) - 2(h) show detailed analyses on the electronic states forming the Fermi surfaces around X point.  Figures \ref{fig2}(c) and 2(f) show the band dispersion of cut1 and cut2 denoted in Fig. \ref{fig2}(a), and corresponding momentum distribution curves (MDCs) are displayed in Figs. \ref{fig2}(d), 2(e) and Figs. 2(g), 2(h), respectively. In order to estimate the peak positions, multi-peak fit using two or four Gaussian functions has been done, and the results are shown in Figs. \ref{fig2}(e) and (h) 
(in the present system with strong atomic disorder \cite{Paris2014}, the MDC width may have rather Gaussian character). 
The Gaussian functions giving the best fit to the bottommost MDCs are displayed in the figures. The peak positions identified by this fit are denoted by circles in Figs. \ref{fig2}(c), 2(d), 2(f), and 2(g) and by vertical bars in Figs. \ref{fig2}(e) and 2(h). The results clearly indicate that there exist two electron-like Fermi surfaces around X point. Full-width-half-maximum was fixed for all the fits. The Fermi wave numbers are estimated as $k_{F1}\sim0.1 $\AA$^{-1}$, $k_{F2}\sim0.12 $\AA$^{-1}$, and $k_{F3}\sim0.22 $\AA$^{-1}$. These labels are indicated in the Figs. 2(e) and 2(h). Identifying the Fermi wave numbers enables us to estimate the approximate amount of electron in the system; the inner and outer electron pockets respectively enclose 3.9\% and 7.2\% of the Brillouin zone. Counting the Luttinger volume of the Fermi surfaces, we estimate the amount of electron to be 0.22 electron per Bi in CeO$_{0.5}$F$_{0.5}$BiS$_2$, that is much smaller than the nominal value. 
Moreover, the topology of the Fermi surface at $x\sim$0.5 predicted by LDA calculation is different from our results, and likely to be closer to the Fermi surface at $x\sim$0.25.

According to the scanning tunneling microscopy study, Bi defects have been found on the BiS plane, whose amount was approximately 3\% of Bi site \cite{Machida2014}. Since one Bi$^{3+}$ defect introduces three holes to the BiS$_2$ layer, the defect of 3\% can eliminate 0.09 electrons per Bi. However, this is not enough to explain the reduction to $x = 0.22$ from $x = 0.5$. One possibility is that the F ions remain trapped in interstitial sites and the amount of the doped electrons is reduced from the nominal F content. The electron probe microanalysis has found that the actual F concentration can deviate largely from the nominal value in this system \cite{Nagao2014}. Therefore, further investigations are needed for the precise F characterization.



\begin{figure}
\includegraphics[width=8.5cm]{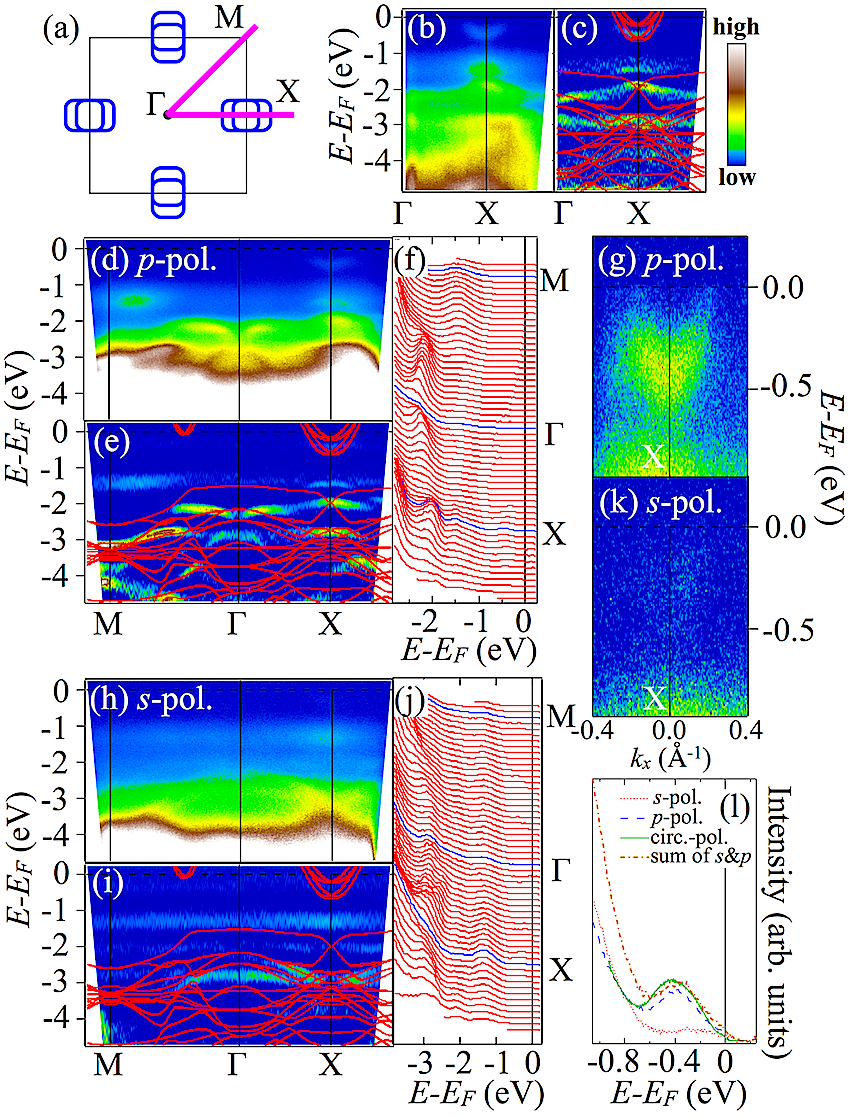}
\caption{(Color online) The polarization dependence of the band dispersions of CeO$_{0.5}$F$_{0.5}$BiS$_2$. (a) Schematic Fermi surface and the cutting line along M-$\Gamma$-X. (b) Raw data and (c) EDC curvature along $\Gamma$-X with circularly polarized light. (d) Raw data, (e) EDC curvature, (f) EDCs along M-$\Gamma$-X, and (g) detailed dispersion around X along $\Gamma$-X with $p$-polarized light in the $k_x$ direction.  (h) Raw data, (i) EDC curvature, (j) EDCs along M-$\Gamma$-X, and (k) detailed dispersion around X along $\Gamma$-X with $s$-polarized light in the $k_x$ direction. The \textit{ab initio} calculation for the  CeO$_{0.5}$F$_{0.5}$BiS$_2$ system is overlaid on each curvature as well. (l) Comparison of EDCs of the electron-like band around X with $p$-, $s$-, circular, and the sum of $s$ and $p$ polarization integrated in the range of $k_x=\pm$0.4 \AA$^{-1}$ along $\Gamma$-X direction.
}
\label{fig3}
\end{figure}

\begin{figure}
\includegraphics[width=7.5cm]{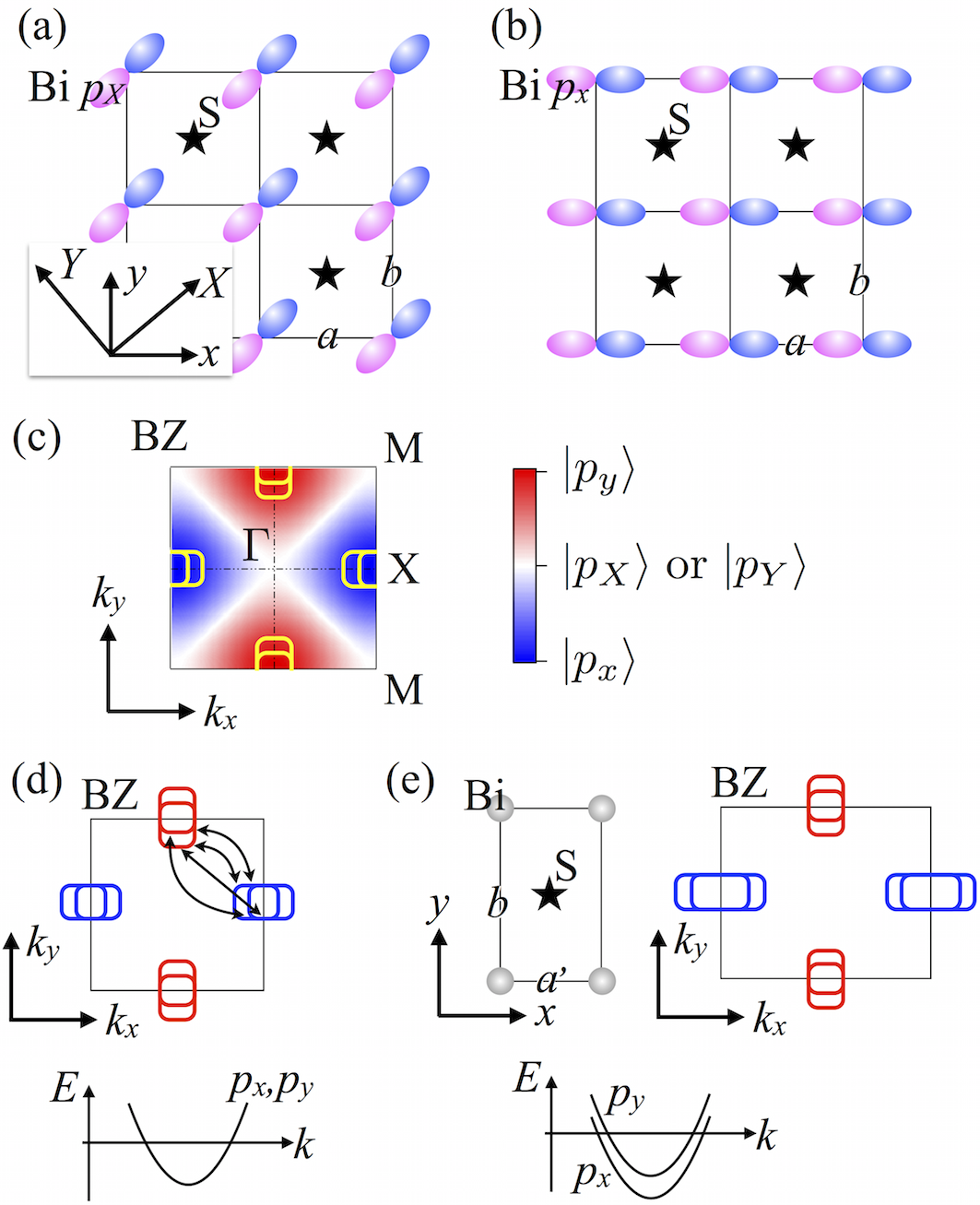}
\caption{(Color online)  
(a) Schematic diagrams of the orbital polarization along $\Gamma$-M predicted by Usui, Suzuki, and Kuroki \cite{Usui2012} and (b) observed orbital polarization along $\Gamma$-X. They can be transformed back and forth each other by their linear combination. (c) A schematic diagram of the orbital distribution proposed by Usui, Suzuki, and Kuroki \cite{Usui2012} and the observed  Fermi surfaces.  (d) Schematic diagrams of the Fermi surface and corresponding band dispersion of Bi 6$p_x$/6$p_y$. Nesting vectors are also shown here. (e) Schematic diagrams of orthorhombically distorted Bi sites with $a'<a$, Fermi surface, and band Jahn-Teller effect on Bi 6$p_x$/6$p_y$.
 }
\label{fig4}
\end{figure}

Figure \ref{fig3}(a) shows the schematic Fermi surface, and we measured ARPES data along M-$\Gamma$-X line as depicted by the bold line. Raw data and EDC curvatures of the valence band (here, EDC stands for energy distribution curve) taken with circularly polarized photons are shown in Figs. \ref{fig3}(b) and 3(c), respectively.
Figures \ref{fig3}(d) - 3(g) and 3(h) - 3(k) show the band dispersions along M-$\Gamma$-X taken with $p$- and  $s$-polarization setup, respectively. Raw data are displayed in Figs. 3(d) and 3(h), corresponding EDC curvatures are shown in Figs. 3(e) and 3(i), EDCs are in Figs. 3(f) and 3(j), and the detailed electron-like band is shown in Figs. 3(g) and 3(k). The \textit{ab initio} calculation is overlaid on each curvature plots. 
One can see a flat band located about  $E=-1.3$ eV (between Bi $6p$ conduction band and valence bands)  due to localized Ce $4f$ electrons for both polarizations in Figs. 3(d) and 3(h) as reported by Ce $4d$-$4f$ resonance experiment on Ce(O,F)BiS$_2$ \cite{Sugimoto2015b}. 
Comparing the band dispersion taken with $s$- and $p$-polarizations, spectral features are clearly observed with $p$-polarization whereas they are hardly seen with $s$-polarization as shown in Figs. \ref{fig3}(d) and 3(h).  The EDCs integrated within $k_x=\pm$0.4 \AA$^{-1}$ along $\Gamma$-X direction are also displayed in Fig. \ref{fig3}(l), and one can see that the spectral weights of $p$- and $s$-polarizations are drastically changed. Considering that the cuts along $\Gamma$-X were taken in the direction of $k_x$, the results suggest us that the electron-like band near X point in the $k_x$ direction (\textit{i.e.}, the Fermi surface is parallel to the $k_y$ direction) is dominantly derived from Bi 6$p_x$ orbital. In other words, the system has quasi-one-dimensional property in at least $\Gamma$-X cut.

According to the minimal model calculation by Usui, Suzuki, and 
 Kuroki \cite{Usui2012}, the band along $\Gamma$-X consists of 
well-hybridized Bi $p_X$ and $p_Y$ (namely, close to pure Bi $p_x$ or 
Bi $p_x$) whereas that along $\Gamma$-M is derived from pure Bi $p_X$ 
or $p_Y$ as in Figs. 4(a) and 4(b). Here, the $X$- and $Y$-axes can be obtained by rotating the 
$x$- and $y$-axes by 45 degrees with respect to $z$-axis. Fig. \ref{fig4}(c) shows the proposed orbital distribution and schematic Fermi surfaces. The basis can be transformed 
by their linear combinations as
\begin{equation}
|p_X\rangle = \frac{1}{\sqrt{2}}\{ |p_x\rangle +|p_y\rangle \}, \;
|p_Y\rangle = \frac{1}{\sqrt{2}}\{ |p_x\rangle -|p_y\rangle \}.
\end{equation}
At the doping level of $x = 0.5$, the band along $\Gamma$-M crosses 
$E_F$ and forms almost strait Fermi surfaces with $p_X$ (or $p_Y$) 
character. Usui, Suzuki, and  Kuroki proposed the Fermi surface 
nesting between the $p_X$ Fermi surfaces or between the $p_Y$ Fermi 
surfaces can enhance the spin fluctuations for the superconductivity 
\cite{Usui2012}. On the other hand, at the doping level of $x$ = 0.25, 
the band along $\Gamma$-M does not cross $E_F$ and only the band along 
the $\Gamma$-X forms Fermi pockets (disconnected Fermi surfaces) 
around X point with almost pure $p_x$ or $p_y$ character.
In this sense, our results are partly consistent with the previous 
calculation for $x$=0.25. Considering the fact that the observed Fermi 
surfaces of the superconducting samples are similar to those predicted 
for $x$=0.25, the mechanism of the superconductivity should be 
discussed based on the disconnected Fermi surfaces. 

Here, we have succeeded in the detailed orbital characterization; in the rectangular 
Fermi surface around $(\pi,0)$, the straight portion parallel to the $k_y$-axis 
is dominated by Bi 6$p_x$ as schematically shown in Fig. \ref{fig4}(d).
On the other hand, the straight portion parallel to the $k_x$-axis 
has more Bi 6$p_y$ character as suggested by the Fermi surface map in Fig. 2(b)
although the orbital selectivity is not exact in this region.
The rectangular Fermi pockets around X point keeps a peculiar 
quasi-one-dimensionality due to the orbital polarization. This 
situation is partly similar to the quasi-one-dimensional Fermi 
surfaces with the orbital polarization for $x = 0.5$ \cite{Usui2012}. 
However, as indicated by the arrows in Fig. 4(d), possible nesting channels 
between the straight portions of the rectangular Fermi pockets may
have different orbital character, indicating spin and orbital fluctuations.
In addition, the Fermi surface nesting may enhance the 
electron-lattice interaction.
The electron-electron interaction (spin and orbital fluctuations) and the 
electron-lattice interaction would be one of the reasons that the band 
dispersions obtained by ARPES showed poor agreements with the {\it ab 
initio} calculation.
Interestingly, the orbitally-polarized rectangular Fermi surfaces can 
give the system a possibility of band Jahn-Teller effect. If the Bi lattice is distorted from tetragonal to orthorhombic, for example, the degeneracy of Bi $p_x$/$p_y$ will be split as 
schematically depicted in Figs. \ref{fig4}(d) and 4(e) (where $a'<a$), resulting 
in the distorted Fermi surface with $C_2$ symmetry. As a result of the 
band Jahn-Teller effect, the system has a better nesting condition than the 
tetragonal symmetry, which further enhances the one-dimensionality.
The one dimensional Fermi surfaces may provide Peierls instability to 
the system. The combination of band Jahn-Teller distortion and Peierls 
instability would be responsible for the monoclinic distortion under 
high pressure.

In summary, we have studied the electronic structure of CeO$_{0.5}$F$_{0.5}$BiS$_2$ by means of polarization dependent ARPES. From the circularly polarized photon, the doped electron has been estimated using Fermi surface as 0.22 electron per Bi in CeO$_{0.5}$F$_{0.5}$BiS$_2$, that is much smaller than the nominal value of 0.5. 
From the linearly polarized photons, we have succeeded in the observation of orbitally polarized Fermi surface. It indicates that the system has quasi-one-dimensional property, that could enhance the electronic correlation and/or electron-phonon coupling. This might be why the band dispersion obtained by ARPES has poor agreements with the \textit{ab initio} calculation. 

We would like to acknowledge Prof. Yoshikazu Mizuguchi, Prof. Yoshihiko Takano, and Prof. Hidetomo Usui for the fruitful discussion. T.S. and D.O. acknowledge the support from  JSPS Research Fellowship for Young Scientists.  This work is partially supported by JSPS KAKENHI (Grant No. 15H03693).  E.F.S. acknowledges financial support from the JSPS postdoctoral fellowship for overseas researchers as well as the Alexander von Humboldt Foundation (Grant No. P13783). The synchrotron radiation experiments have been done with the approval of Hiroshima Synchrotron Radiation Center (Proposal No.14-A-13 and No.14-B-24).

\end{document}